\documentclass[conference, 10pt]{IEEEtran}
\IEEEoverridecommandlockouts
\usepackage{cite}
\usepackage{amsmath,amssymb,amsfonts}
\usepackage{graphicx}
\usepackage{textcomp}
\usepackage{amsmath}
\usepackage{colortbl}
\usepackage{booktabs} 
\usepackage{multirow}
\usepackage{hyperref}       % hyperlinks
\usepackage{url}            % simple URL typesetting
\usepackage{booktabs}       % professional-quality tables
\usepackage{amsfonts}       % blackboard math symbols
\usepackage{nicefrac}       % compact symbols for 1/2, etc.
\usepackage{microtype}      % microtypography
\usepackage{xcolor}         % colors
\usepackage{algorithm}
\usepackage{colortbl}
\usepackage{stfloats}
\usepackage{caption}
\usepackage{algorithmic}
\usepackage{amsmath}
\usepackage{tikz}
\newcommand*\circled[1]{\tikz[baseline=(char.base)]{
            \node[shape=circle,draw,inner sep=2pt] (char) {#1};}}
\usepackage{orcidlink}
\usepackage{comment}
\usepackage{amssymb}
\usepackage{braket}
\usepackage{xcolor}
\usepackage[font=scriptsize,skip=0pt]{caption}

\newcommand{\rpoint}[1]{\scalebox{0.7}{\circled{{\fontfamily{pcr}\selectfont\footnotesize #1}}}}

\usepackage{cleveref}
\usepackage{setspace}
\def\BibTeX{{\rm B\kern-.05em{\sc i\kern-.025em b}\kern-.08em
    T\kern-.1667em\lower.7ex\hbox{E}\kern-.125emX}}

\usepackage{pifont}
\usepackage{enumitem}
\usepackage{hyperref}
\setlist[itemize]{leftmargin=*}%
\setlist[enumerate]{leftmargin=*}%

\usepackage[compact]{titlesec}

\titlespacing\section{0pt}{0.3\baselineskip}{0.2\baselineskip}
\titlespacing\subsection{0pt}{0.2\baselineskip}{0.1\baselineskip}
\titlespacing\subsubsection{0pt}{0.15\baselineskip}{0.1\baselineskip}

\begin{document}

% \bstctlcite{mybstctl}

% \title{QNAS Under BP: Trainability and Expressibility Guided Hybrid Neural Architecture Search}

% \title{Rethinking Expressibility vs Trainability in \\ Hybrid QNNs: Need for Multi-Objective NAS}

\title{Rethinking Expressibility-Trainability Trade-off in Hybrid Quantum Neural Networks }

\author{\IEEEauthorblockN{Muhammad Kashif\orcidlink{0000-0003-2023-6371}\textsuperscript{1,2}, and Muhammad Shafique\orcidlink{0000-0002-2607-8135}\textsuperscript{1,2}\\
 \IEEEauthorblockA{
 \textsuperscript{1}eBRAIN Lab, Division of Engineering, New York University Abu Dhabi (NYUAD), Abu Dhabi, UAE\\
 \textsuperscript{2}Center for Quantum and Topological Systems (CQTS), NYUAD Research Institute, NYUAD, Abu Dhabi, UAE\\
 muhammadkashif@nyu.edu, muhammad.shafique@nyu.edu
}}

% \vspace{-15pt}
}

\maketitle

\begin{abstract}

Hybrid quantum neural networks (HQNNs) integrate parameterized quantum circuits (PQCs) within classical networks, where the behavior of the underlying PQCs is often the primary focus of analysis. In this context, expressibility and trainability are widely used to characterize PQC's performance and are commonly assumed to exhibit a trade-off, where highly expressive circuits are more susceptible to barren plateaus. However, the validity of this relationship in HQNNs remains unclear.
In this paper, we systematically analyze the expressibility--trainability relationship in HQNNs across varying circuit depths, qubit counts, entanglement topologies. We consider different training configurations, including pure PQCs, quantum-only training in hybrid setting, and full end-to-end training of hybrid models. Our results show that pure PQCs exhibit only a weak and regime-dependent trade-off, while hybrid architectures increasingly disrupt and can eliminate this relationship under full hybrid training. This indicates that classical components reshape the optimization landscape, decoupling trainability from PQC expressibility.
We further propose a multi-objective neural architecture search (NAS) framework that jointly optimizes expressibility, trainability, and task performance over a combined classical--quantum design space, revealing different Pareto-optimal solutions under full end-to-end and quantum only training in hybrid setting. different trainability definitions. Our results suggest that hybridization is not just an implementation detail, but a defining factor in the performance of quantum machine learning models.

\end{abstract}
\begin{IEEEkeywords}
Quantum machine learning, Quantum neural networks, Hybrid quantum neural networks, Expressibility, Trainability

\end{IEEEkeywords}
%
% \begin{spacing}{0.95}
%
\section{Introduction}
Variational quantum algorithms (VQAs) are widely studied for leveraging noisy intermediate-scale quantum (NISQ) devices~\cite{cerezo2021variational}. At the core of these methods are parameterized quantum circuits (PQCs), which act as trainable models in applications ranging from optimization to quantum machine learning~\cite{amaro2022filtering,innan2025quav,kashif2025position,schuld2015introduction,kashif2025computational}. 
To this end, the effectiveness of PQCs is commonly characterized by two key properties: \textit{(i)} expressibility, which quantifies the ability of a circuit to represent complex quantum states~\cite{Sim_2019}, and \textit{(ii)} trainability, which reflects the ease of optimizing circuit parameters using gradient-based methods~\cite{heyraud2023efficient}.
A widely observed phenomenon in PQCs is the expressibility--trainability trade-off. As circuits become more expressive (approaching approximate unitary $2$-designs),their gradients tend to vanish exponentially with system size, leading to barren plateaus~\cite{mcclean2018barren,kashif2024alleviating,Holmes_2022,kashif2024resqnets}. This trade-off has motivated extensive efforts to balance circuit depth, entanglement, and parameterization for achieving both sufficient expressibility and practical trainability~\cite{tangpanitanon2020,roseler2026find,maronese2026high,kashif2023unified,sanz2026efficiently,yu:2025,thanasilp2023}.

\subsection{Hybrid Quantum–Classical Models: An Open Question}
Hybrid quantum--classical neural networks (HQNNs) have emerged as a practical framework in quantum machine learning (QML), where PQCs are embedded within classical networks and combined with classical feature extraction and post-processing layers~\cite{kashif2021design,paquet2022quantumleap,sagingalieva2023hybrid,liu2021hybrid,kashif2026design}. 
While the behavior of PQCs is often the primary focus of analysis in such models, the inclusion of classical components can significantly reshape the overall optimization landscape. As a result, it remains unclear whether the expressibility--trainability trade-off observed in standalone PQCs continues to hold in hybrid settings.
This leads to the central question of this work:
\textit{\textbf{Does the expressibility--trainability relationship persist in HQNNs, and can PQC design principles derived from standalone settings be directly applied to hybrid architectures?}}
Despite extensive studies on PQCs, this question has not been systematically addressed. Existing works either focus on standalone circuits or implicitly assume that their properties transfer to hybrid models, leaving the validity of the trade-off in HQNNs largely unexplored.

\subsection{Design Implications for Hybrid Models }
The answer to the above-mentioned question has direct implications for the design of HQNNs. In standalone PQCs, the expressibility--trainability trade-off serves as a key guideline for architecture design. If this relationship does not hold in hybrid models, such heuristics may no longer be reliable.
This raises a fundamental challenge:
\textit{\textbf{In the absence of a predictable relationship between expressibility and trainability, how should HQNNs be designed?}}
Without clear design principles, manual architecture selection becomes increasingly difficult, particularly as hybrid models involve complex interactions between classical and quantum components. This motivates the need for systematic and automated approaches such as neural architecture search (NAS)~\cite{kashif2025faqnas,marchisio2026hybrid,kashif2026closing} that can jointly account for multiple objectives, including expressibility, trainability, and task performance.

\subsection{Our Novel Contributions}

We address the above questions through a systematic study of the expressibility--trainability relationship in HQNNs. Our main contributions are:

\begin{itemize}

    \item We conduct a systematic analysis of expressibility and trainability in HQNNs across different settings, varying circuit depth, qubit count, and entanglement topology, and compare standalone PQCs with hybrid models under different training configurations, including quantum-only and fully trainable hybrid settings (\textbf{Section~\ref{sec:meth_phase1}}).

    \item We show that, while standalone PQCs exhibit a weak and regime-dependent expressibility--trainability trade-off, this relationship becomes inconsistent in hybrid architectures and can be significantly weakened or eliminated under full hybrid training (\textbf{Section~\ref{sec:res_phase1}}).

    \item We demonstrate that different definitions of trainability based on \textit{(i)} full hybrid model gradients and \textit{(ii)} quantum-only gradients in hybrid models, lead to decoupling between expressibility and trainability, revealing an inherent ambiguity in assessing optimization behavior in HQNNs (\textbf{Section~\ref{sec:res_phase2}}).

    \item Motivated by the absence of reliable design principles, we propose a multi-objective neural architecture search (NAS) framework that jointly optimizes expressibility, trainability, and task performance over a combined classical--quantum design space, and analyze the resulting Pareto-optimal architectures (\textbf{Section~\ref{sec:meth_phase2}}).

    \item We show that different notions of trainability yield distinct Pareto fronts, highlighting that effective HQNN design requires balancing competing objectives from both classical and quantum design space rather than optimizing a single metric (\textbf{Section~\ref{sec:res_phase2}}).

\end{itemize}
%%%%%%%%%%%%%%%%%%%%%%%%%%%%%%%%%%%%%%%%%%%%%%%%%%%%%%%%%%%%%%%%%%%%%%%%%%%%%%%%%%%%%%%%%%%%%%%%%%%%%%%%%%%%%%%%%%%%%%%%%%%%%%%%%%%%%%%%%%%%%%%
\section{Background \& Preliminaries}

\subsection{Parameterized Quantum Circuits}

VQAs are based on PQCs of the form~\cite{cerezo2021variational}:

\begin{equation}
U(\boldsymbol{\theta}) = \prod_{l=1}^{L} U_l(\boldsymbol{\theta}_l)
\end{equation}

where $\boldsymbol{\theta} \in \mathbb{R}^p$ denotes the set of trainable parameters and $L$ is the circuit depth. Acting on an initial state $|0\rangle^{\otimes n}$, the circuit prepares:
\begin{equation}
|\psi(\boldsymbol{\theta})\rangle = U(\boldsymbol{\theta}) |0\rangle^{\otimes n}
\end{equation}

For a given observable $O$, the model output is defined as:
\begin{equation}
f(\boldsymbol{\theta}) = \langle \psi(\boldsymbol{\theta}) | O | \psi(\boldsymbol{\theta}) \rangle
\end{equation}

\subsection{Hybrid Quantum--Classical Neural Networks}

In hybrid quantum-classical neural networks (HQNNs), PQCs are embedded within classical architectures~\cite{kashif2023impact}. A typical hybrid model can be expressed as:
\begin{equation}
y = g_{\boldsymbol{\phi}}\big(f_{\boldsymbol{\theta}}(h_{\boldsymbol{\phi}}(x))\big)
\end{equation}
where $h_{\boldsymbol{\phi}}$ denotes a classical feature mapping, $f_{\boldsymbol{\theta}}$ represents the PQC, and $g_{\boldsymbol{\phi}}$ is a classical post-processing function. The parameters $\boldsymbol{\phi}$ and $\boldsymbol{\theta}$ correspond to classical and quantum components, respectively.

\subsection{Expressibility of VQAs}

Expressibility measures the ability of a parameterized quantum circuit to represent a wide range of quantum states. It is commonly characterized by comparing the distribution of states generated by the circuit to that of Haar-random states over the Hilbert space.
A standard approach is to compute the Kullback-Leibler (KL) divergence between the fidelity distribution induced by the circuit, $P_{\text{PQC}}(F)$, and the corresponding Haar-random fidelity distribution, $P_{\text{Haar}}(F)$~\cite{Sim_2019}:
\begin{equation}
\mathcal{E} = D_{\mathrm{KL}}\big(P_{\text{PQC}}(F) \,\|\, P_{\text{Haar}}(F)\big)
\end{equation}

Lower values of $\mathcal{E}$ indicate that the circuit more closely approximates Haar-random states, and is therefore more expressive, while higher values indicate limited coverage of the Hilbert space.
While this definition provides a general characterization of circuit expressibility, alternative task-driven proxies can be used depending on the application, we discuss this more when we discuss trainability in our methodology (Section~\ref{sec:expr_est_phase1}).

\subsection{Trainability of VQAs}

Trainability of PQCs is commonly characterized by the variance of gradients of a cost function with respect to circuit parameters. Given a cost function $C(\boldsymbol{\theta})$, trainability can be quantified using the variance of gradients~\cite{mcclean2018barren}:

\begin{equation}\label{eq:gradvar_genral}
\mathrm{Var}\left[\frac{\partial C}{\partial \theta_k}\right]
\end{equation}

In highly expressive PQCs, this variance can vanish exponentially with the number of qubits, leading to barren plateaus and making optimization difficult~\cite{mcclean2018barren,Holmes_2022}.
Gradients of quantum circuits are typically computed using the parameter-shift rule~\cite{wierichs2022general}. For a parameter $\theta_k$ associated with a parametrized gate ($Rx(\theta)$,$Ry(\theta)$,$Rz(\theta)$), the gradient can be expressed as:
\begin{equation}\label{eq:param_shift}
\frac{\partial C}{\partial \theta_k}
=
\frac{1}{2}
\left[
C\left(\boldsymbol{\theta} + \frac{\pi}{2}\mathbf{e}_k\right)
-
C\left(\boldsymbol{\theta} - \frac{\pi}{2}\mathbf{e}_k\right)
\right]
\end{equation}
where $\mathbf{e}_k$ is the unit vector along the $k$-th parameter direction.

% This formulation enables exact gradient evaluation using repeated circuit executions.

%%%%%%%%%%%%%%%%%%%%%%%%%%%%%%%%%%%%%%%%%%%%%%%%%%%%%%%%%%%%%%%%%%%%%%%%%%%%%%%%%%%%%%%%%%%%%%%%%%%%%%%%%%%%%%%%%%%%%%%%%%%%%%%%%%%%%%%%%%%%%%%%

\section{Our Methodology}\label{sec:methodology}

\begin{figure*}
    \centering
    \includegraphics[width=1.0\linewidth]{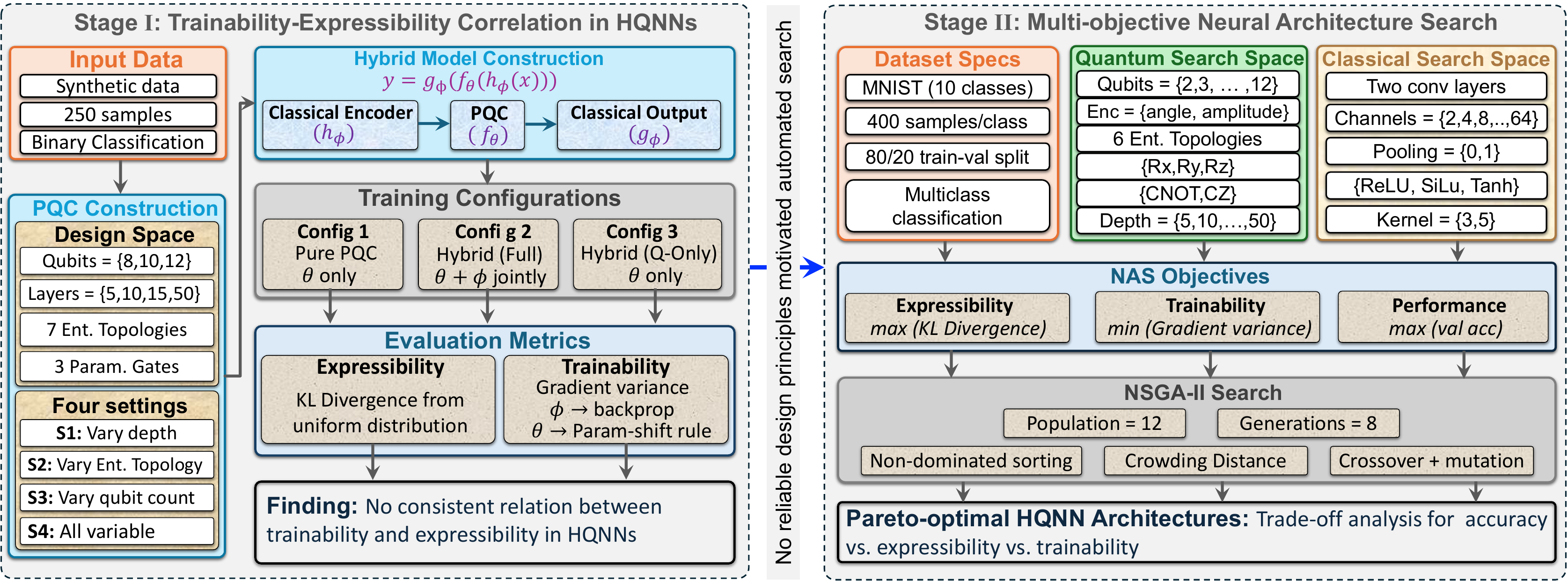}
    \caption{Overview of the our two-stage methodology. \textbf{Stage I} presents the empirical analysis of the expressibility-trainability relationship in HQNNs. A flexible PQC design space is explored under multiple controlled settings and training configurations, including pure PQCs, full hybrid training, and quantum-only training. Expressibility (via KL divergence) and trainability (via gradient variance) are evaluated, revealing no consistent relationship between the two in hybrid models. \textbf{Stage II} introduces a multi-objective NAS framework over a joint classical-quantum design space. Architectures are optimized with respect to expressibility, trainability, and task performance using NSGA-II, resulting in Pareto-optimal HQNN designs that capture trade-offs between accuracy, expressibility and trainability.}
    \label{fig:meth_1}
\end{figure*}

Our study follows a two-stage approach. In the first stage, we empirically analyze the relationship between expressibility and trainability in HQNNs. In the second stage, we formulate HQNN design as a multi-objective optimization problem and address it using neural architecture search (NAS). An overview of our methodology is presented in Fig.~\ref{fig:meth_1}.

\subsection{Stage I: Empirical Analysis of Expressibility-Trainability Trade-off in HQNNs} \label{sec:meth_phase1}

\subsubsection{Input Data and Task Setup}\label{sec:datasettask_phase1}

\begin{figure}
    \centering
    \includegraphics[width=1.0\linewidth]{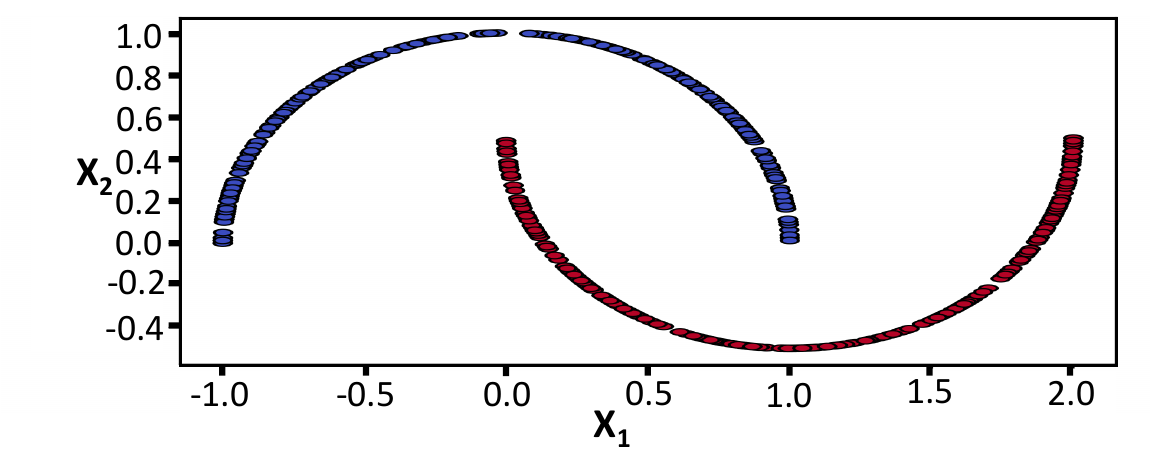}
    \caption{Synthetic dataset used for Expressibility-Trainability analysis of HQNNs}
    \label{fig:dataset_s1}
\end{figure}

For a controlled and unified input distribution for evaluating trainability, we use a synthetic two-class dataset consisting of nonlinearly separable samples in a low-dimensional feature space, as shown in Fig.~\ref{fig:dataset_s1}. The dataset contains $500$ samples, split into training and validation sets using a $70/30$ split.
The task is formulated as a binary classification problem with cross-entropy loss:
\begin{equation}
\mathcal{L} = - \sum_{c=1}^{2} y_c \log \hat{y}_c
\end{equation}
where $y_c$ denotes the ground-truth label and $\hat{y}_c$ represents the predicted class probability.
Unlike conventional learning setups, our objective is not to optimize predictive performance, but to analyze intrinsic properties of PQCs in both standalone and hybrid settings. To this end, we intentionally choose a simple dataset that serves primarily for task-dependent loss function definition, and provide a consistent input distribution for gradient evaluation. Since the trainability metric (Eq.~\ref{eq:trainability_phase1}) is based on the variance of parameter gradients, maintaining a controlled input distribution is essential for obtaining meaningful and reliable estimates of the optimization landscape. This consideration is particularly important in hybrid architectures, where classical preprocessing layers transform the input before quantum encoding, thereby directly influencing the resulting gradient behavior.

\subsubsection{PQC Construction} \label{sec:PQC_const_stage1}

\begin{figure}
    \centering
    \includegraphics[width=1.0\linewidth]{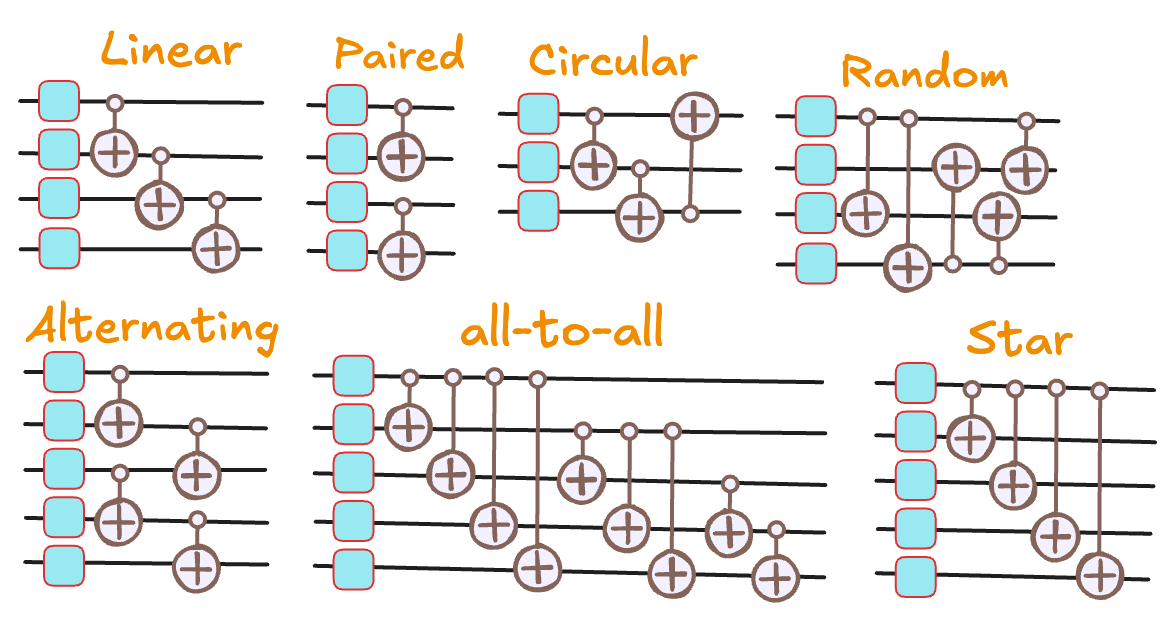}
    \caption{Entanglement topologies used in this paper}
    \label{fig:ent_topo}
\end{figure}

We consider a flexible design space of PQCs rather than a single fixed architecture, as expressibility and trainability are highly dependent on circuit structure, including depth and entanglement patterns. Analyzing a single architecture may lead to conclusions that are specific to that design and do not generalize.
The overall design space, we used in this paper is defined as:
\[
\mathcal{A} = (n, L, \mathcal{P}, \mathcal{T})
\]
where $n$ denotes the number of qubits, $L$ the circuit depth, $\mathcal{P}$ the set of parameterized single-qubit rotation gates, and $\mathcal{S}$ the entanglement topology.
In our experiments, architectures are generated by randomly sampling from this space. Specifically, $n \in \{8,10,12\}$, $L \in \{5,10,15,20\}$, and rotation gates are randomly selected from $\mathcal{P} = \{R_X, R_Y, R_Z\}$, with one gate assigned per qubit. The entanglement topology is sampled from
$\mathcal{S}~\in~\{\text{linear}, \text{paired}, \text{circular}, \text{random}, \text{alternating}, \text{all-to-all}, \text{star}\}$, and CNOT gate is used for qubit entanglement. The entanglement topologies used are shown in Fig.~\ref{fig:ent_topo}. The random topology is illustrated using a representative instance sampled from the space of possible connectivity patterns, and may vary across different realizations.
To systematically analyze the effect of different architectural factors, we consider multiple experimental settings, summarized in Table~\ref{tab:pqc_settings}.
This design enables us to isolate the impact of individual architectural factors, as well as their combined effects on expressibility and trainability.

\begin{table}[h]
\centering
\footnotesize
\caption{Different training configurations for PQC construction in Stage~I}
\label{tab:pqc_settings}
\resizebox{\columnwidth}{!}{
\begin{tabular}{lccc}
\toprule
\rowcolor{gray!15}
\textbf{Setting} & \textbf{Qubits ($n$)} & \textbf{Depth ($L$)} & \textbf{Topology ($\mathcal{S}$)} \\
\midrule
\rowcolor{gray!5}
Setting 1 & Variable & Fixed ($10$) & Fixed (alternating) \\
\rowcolor{gray!5}
Setting 2 & Fixed ($10$) & Fixed ($10$) & Variable \\
\rowcolor{gray!5}
Setting 3 & Fixed ($10$) & Variable & Fixed (alternating) \\
\rowcolor{gray!5}
Setting 4 & Variable & Variable & Variable \\
\bottomrule
\end{tabular}}
\end{table}
%
% \begin{itemize}
%     \item \textbf{Setting 1:} Fixed number of qubits ($n=10$), and fixed entanglement topology (alternating), with only circuit depth varied randomly from $L$. 
    
%     \item \textbf{Setting 2:} Fixed number of qubits ($n=10$), and circuit depth ($L=10$), and entanglement topology sampled uniformly at random from $\mathcal{S}$.
    
%     \item \textbf{Setting 3:} Fixed circuit depth ($L=10$), and entanglement topology (alternating), while number of qubits are sampled uniformly at random from $n$.
    
%     \item \textbf{Setting 4:} All parameters are randomly sampled.
% \end{itemize}
%

\subsubsection{Hybrid Model Construction}

For each sampled PQC, we construct a hybrid quantum-classical neural network (HQNN) of the form:
\begin{equation}\label{eq:hyb_model}
    y = g_{\boldsymbol{\phi}}\big(f_{\boldsymbol{\theta}}(h_{\boldsymbol{\phi}}(x))\big)
\end{equation}
where $h_{\boldsymbol{\phi}}$ is a classical feature extractor, $f_{\boldsymbol{\theta}}$ is the PQC, and $g_{\boldsymbol{\phi}}$ is a classical output layer.
The classical feature extractor $h_{\boldsymbol{\phi}}$ is implemented as a fully connected layer followed by a $\tanh$ activation, mapping the 2-dimensional input into a latent representation compatible with the number of qubits. The PQC processes this representation using angle encoding, where features are mapped to rotation angles via $R_Y$ gates. The circuit outputs expectation values of Pauli-$Z$ operators, one per qubit, which are then passed to a final linear layer $g_{\boldsymbol{\phi}}$ producing logits for binary classification. By embedding the same (standalone) PQC architectures within this hybrid structure, we can directly compare their behavior under pure quantum and hybrid settings, and assess how classical-quantum interactions affect the relationship between expressibility and trainability. For optimization, we use the Adam optimizer with a learning rate of $0.01$. The model is trained for $15$ epochs.

\subsubsection{Training Configurations} \label{sec:training_configurations}

To analyze how the relationship between expressibility and trainability evolves under different training setups, we consider three configurations:

\begin{itemize}

    \item \textbf{\textit{Configuration 1 - Pure PQC:}} 
    The PQC is trained as a standalone model without classical pre- or post-processing layers. The loss is computed directly from the quantum outputs $f_{\boldsymbol{\theta}}(x)$, corresponding to the quantum component of Eq.~\ref{eq:hyb_model}, and only quantum parameters $\boldsymbol{\theta}$ are optimized. This configuration serves as a baseline to characterize the intrinsic behavior of isolated quantum circuits.

    \item \textbf{C\textit{onfiguration 2 - Hybrid Model (Full Training):}} 
    The PQC is embedded within a hybrid architecture with classical pre- and post-processing layers. Both classical parameters $\boldsymbol{\phi}$ and quantum parameters $\boldsymbol{\theta}$ in Eq.~\ref{eq:hyb_model} are optimized jointly in an end-to-end manner. Classical components participate in both forward and backward passes.

    \item \textbf{\textit{Configuration 3 - Hybrid Model (Quantum-Only Training):}} 
    The PQC is evaluated within the hybrid architecture, but only quantum parameters $\boldsymbol{\theta}$ (Eq.~\ref{eq:hyb_model}) are updated, while classical parameters $\boldsymbol{\phi}$ (Eq.~\ref{eq:hyb_model}) remain fixed. This isolates the behavior of the quantum circuit in the presence of classical transformations, without adaptation from classical layers.

\end{itemize}

Each sampled architecture is evaluated under all three configurations for every architectural setting (discussed in Section~\ref{sec:PQC_const_stage1}, Table~\ref{tab:pqc_settings}). This enables a consistent comparison between the intrinsic behavior of PQCs and their behavior within hybrid architectures, allowing us to assess how classical components influence the expressibility--trainability relationship.

\subsubsection{Trainability Analysis}\label{sec:train_anal_phase1}

Trainability is quantified using the variance of cost-function gradients. Since HQNNs contain both quantum parameters $\boldsymbol{\theta}$ and classical parameters $\boldsymbol{\phi}$, we consider two notions of trainability.
For both standalone PQC (Configuration 1) and quantum-only trainability in hybrid setting (Configuration 2), we compute the gradient variance with respect to the quantum parameters:
\[\mathrm{Var}\left[\nabla_{\boldsymbol{\theta}} C\right]
\]
For full-model trainability, we compute the gradient variance over all trainable parameters:
\[
\mathrm{Var}\left[\nabla_{\boldsymbol{\omega}} C\right],
\qquad
\boldsymbol{\omega} = \{\boldsymbol{\theta},\boldsymbol{\phi}\}
\]
Quantum gradients are computed using the parameter-shift rule (Eq.~\ref{eq:param_shift}), while classical gradients are obtained using standard backpropagation. For each architecture and training configuration, gradients are evaluated over a mini-batch of $100$ training samples, and the variance is computed over the resulting gradient components.
The trainability metric is then defined as:
\begin{equation}\label{eq:trainability_phase1}
   \mathcal{T}_{\boldsymbol{\alpha}} = -\log_{10}
   \left(
   \mathrm{Var}\left[\nabla_{\boldsymbol{\alpha}} C\right]
   \right),
   \qquad
   \boldsymbol{\alpha} \in \{\boldsymbol{\theta}, \boldsymbol{\omega}\}
\end{equation}
where $\boldsymbol{\alpha}=\boldsymbol{\theta}$ corresponds to quantum-only trainability, and $\boldsymbol{\alpha}=\boldsymbol{\omega}$ corresponds to full-model trainability. Lower values of $\mathcal{T}_{\boldsymbol{\alpha}}$ indicate better trainability.

\subsubsection{Expressibility Analysis} \label{sec:expr_est_phase1}

Unlike standard fidelity-based definitions of expressibility, which quantify the coverage of the Hilbert space~\cite{Sim_2019}, we focus on a task-relevant notion of expressibility based on the output probability distribution. This is particularly important in practical learning settings, where measurement outcomes directly influence downstream tasks.
This perspective is further motivated by prior work linking uniform exploration of the Hilbert space to improved performance in classification tasks~\cite{hubregtsen2021evaluation,zhang2025learning}.
Accordingly, we quantify expressibility using a KL divergence-based measure that evaluates the uniformity of the output probability distribution.
For each architecture, we sample random parameter values from $\boldsymbol{\theta} \sim [0, 2\pi)$ and draw input features $x$ from the dataset. For each pair $(\boldsymbol{\theta}, x)$, we compute the corresponding quantum state. Measurement probabilities over computational basis states are obtained as:
\[
p_i = |\langle i \mid \psi(\boldsymbol{\theta}) \rangle|^2
\]

Expressibility is defined as the normalized Kullback--Leibler (KL) divergence between the resulting distribution and the uniform distribution:
\begin{equation} \label{eq:expr_phase1}
    \mathcal{E} = \frac{1}{\log d} \sum_{i} p_i \log \frac{p_i}{1/d}
\end{equation}
where $d = 2^n$ is the Hilbert space dimension.
The metric $\mathcal{E}$ is averaged over $200$ random parameter samples for each architecture to ensure robustness. Lower values of $\mathcal{E}$ indicate distributions closer to uniform and, therefore, higher expressibility.

\subsubsection{Expressibility--Trainability Relationship Analysis}

We analyze the expressibility--trainability relationship by: 
\textit{(i)} evaluating trainability $\mathcal{T}$ (Eq.~\ref{eq:trainability_phase1}) and expressibility $\mathcal{E}$ (Eq.~\ref{eq:expr_phase1}), across a diverse set of architectures under different design settings (Section~\ref{sec:PQC_const_stage1}); 
\textit{(ii)} comparing trends across the three training configurations (Pure PQC, Hybrid Full Training, and Hybrid Quantum-Only Training); 
\textit{(iii)} examining the relationship between $\mathcal{E}$ and $\mathcal{T}$; and 
\textit{(iv)} analyzing classification accuracy to understand how these properties relate to downstream task performance.
This framework enables us to assess whether the commonly assumed expressibility--trainability trade-off persists in HQNNs and to quantify its impact on practical learning performance.

%%%%%%%%%%%%%%%%%%%%%%%%%%%%%%%%%%%%%%%%%%%%%%%%%%%%%%%%%%%%%%%%%%%%%%%%%%%%%%%%%%%%%%%%%%%%%%%%%%%%%%%%%%%%%%%%%%%%%%%%%%%%%%%%%%%%%%%%%%%%%%%%%%%%%%%%%%%%%%%%%%%%%%%%

\subsection{Stage II: Multi-Objective Neural Architecture Search for HQNNs}\label{sec:meth_phase2}
Based on the analysis conducted in Stage~I (results presented in Section~\ref{sec:res_phase1}), we observe that unlike standalone PQCs, no consistent relationship exists between expressibility and trainability in HQNNs. 
In the context of standalone PQCs, there exist an approximate inverse relationship between these properties, which has guided the design of various ansatz families with different expressibility--trainability characteristics.
The absence of such a guiding principle in HQNNs suggests that manual architecture design may lead to suboptimal or biased configurations. This motivates the use of Neural Architecture Search (NAS) to systematically explore architectures that jointly optimize expressibility, trainability, and task performance.

\subsubsection{Problem Formulation}

We formulate HQNN design as a multi-objective optimization problem over architectures $\mathcal{A}$:
\[
\max_{\mathcal{A}} \; \mathrm{Accuracy}(\mathcal{A}), \quad
\min_{\mathcal{A}} \; \mathcal{E}(\mathcal{A}), \quad
\min_{\mathcal{A}} \; \mathcal{T}(\mathcal{A}),
\]
where $\mathcal{E}$ denotes expressibility and $\mathcal{T}$ denotes trainability.
A key distinction in HQNNs is that, while expressibility is primarily an intrinsic property of the PQC, trainability is influenced by both quantum and classical components. In particular, classical layers can modify the effective input representation and gradient flow, thereby reshaping the optimization landscape of the hybrid model.
As a result, optimizing the PQC in isolation may lead to incomplete or misleading conclusions. To address this, we perform NAS to systematically explore the joint classical--quantum design space, where both components collectively determine representational capacity, optimization behavior, and task performance.

\subsubsection{Dataset Specifications}

We evaluate candidate architectures on a multi-class image classification task using a subset of the MNIST image dataset. Specifically, we select all $10$ classes with $400$ samples per class, resulting in a total of $4000$ samples. The dataset is split into training and validation sets using an $80$/$20$ split.
Images are normalized and used as inputs to the hybrid model. This setup provides a sufficiently complex yet computationally efficient benchmark, enabling the evaluation of a large number of candidate architectures during the search process.

\subsubsection{Quantum and Classical Search Space}

Each candidate architecture consists of a classical convolutional preprocessing layers followed by a PQC and a fixed classical output layer. The search space therefore includes both classical and quantum design variables.

\paragraph{Quantum Search Space}
We define the search space for PQCs as:

\[
\mathcal{A}_{q} = (\mathrm{E}_{\mathrm{enc}, n, L, \mathcal{P}, \mathcal{S}, \mathcal{U} })
\]

where $\mathrm{E}_{\mathrm{enc}}$ is the encoding scheme, $n$ is the number of qubits, $L$ is the circuit depth, $\mathcal{P}$ is the set of parameterized rotation gates, $\mathcal{S}$ is the entanglement topology, $\mathcal{U}$ is the entangling gate.
During the search, the classical-to-quantum data encoding is selected from $\{\text{angle-}X, \text{angle-}Y, \text{angle-}Z\}$, $n~\in~\{2,3,\ldots,12\}$, $L~\in\{5,10,15,20,25,30,35,45,50\}$, $\mathcal{P} = \{R_X, R_Y, R_Z\}$, where each qubit is assigned a parameterized rotation gate independently. The entanglement topology is sampled from $\mathcal{S} \in \{\text{linear}, \text{paired}, \text{circular}, \text{alternating}, \text{all-to-all}, \text{star}\}$ (Fig.~\ref{fig:ent_topo}), and the entangling gate is selected from $\mathcal{U} \in \{\mathrm{CNOT}, \mathrm{CZ}$\}.

\paragraph{Classical Search Space}
We use $2$ pre-processing convolutional layers. For each convolutional layer, the search includes:~$\text{channels}~\in~\{2,4,8,12,16,24,32,48,64\}$, $\text{kernel size}~\in~\{3,5\}$, and a binary pooling choice. The activation function is selected from~$\{\mathrm{ReLU}, \mathrm{SiLU}, \tanh\}$.

Based on the search space definitions, each candidate architecture jointly specifies the classical preprocessing  convolutional layers and PQC structure, followed by a fixed classical output layer with $10$ neurons corresponding to the number of classes in MNIST dataset we used for NAS.

\subsubsection{The Architecture Search Procedure}

We aim to identify Pareto-optimal HQNN architectures under three objectives.% $\max \mathrm{Acc}, \min \mathcal{E}, \min \mathcal{T}$.

\paragraph{Objective Functions}\label{sec:NAS_objectives}

We optimize three objectives: trainability, expressibility and task performance. 
We consider two variants of the trainability objective in NAS, \textit{(i)} using gradients with respect to all trainable parameters (Configuration~2 in Section~\ref{sec:training_configurations}), and \textit{(ii)} using gradients with respect to quantum parameters only (Configuration~3 in Section~\ref{sec:training_configurations}). The trainability metric follows the same definition as in Section~\ref{sec:train_anal_phase1}, Eq.~\ref{eq:trainability_phase1}. Since we take $-log_{10}$ of gradient variance, where lower values indicates better trainability, this objective is minimized during search.  
Expressibility is evaluated using the KL divergence-based proxy, discussed in Section~\ref{sec:expr_est_phase1}. This objective is also minimized during the search.
Task performance is measured using validation accuracy (Eq.~\ref{eq:task_performance}), and is maximized as part of the multi-objective optimization.

\begin{equation} \label{eq:task_performance}
    \mathrm{Acc} = \frac{1}{N_{\mathrm{val}}}\sum_{i=1}^{N_{\mathrm{val}}} \mathbb{I}(\hat{y}_i = y_i)
\end{equation}

The optimization is performed separately for two trainability variants: full-model and quantum-only trainability. In both cases, NSGA-II minimizes expressibility $\mathcal{E}$ and trainability $\mathcal{T}$, while maximizing validation accuracy. For consistency with the minimization framework, accuracy is reformulated as $1 - \mathrm{Accuracy}$.
The output of NSGA-II is a set of non-dominated architectures forming the Pareto front. An architecture is Pareto-optimal if no other candidate improves all objectives simultaneously. Analysis of the Pareto front provides insights into how different architectural choices balance accuracy, expressibility, and trainability.

\paragraph{NSGA-II Workflow}
The search space is combinatorial and grows rapidly with the number of design choices across quantum and classical components, making exhaustive exploration infeasible. We therefore use NSGA-II genetic algorithm to search for Pareto-optimal HQNN architectures under the three objectives defined above. Each individual represents a candidate architecture sampled from the joint classical--quantum design space.
To enable a tractable yet representative search, we use a population size of $12$ evolved over $8$ generations, resulting in the evaluation of approximately $108$ candidate architecture, including both the initial population and generated offspring..
Each candidate is trained for $10$ epochs before objective evaluation, balancing computational cost with reliable estimation of performance, expressibility, and trainability.
An overview of NSGA-II workflow is presented in Algorithm~\ref{alg:nsga_hqnn}.

\begin{algorithm}
\caption{NSGA-II-based NAS for HQNNs}
\label{alg:nsga_hqnn}
\footnotesize
\begin{algorithmic}[1]
\STATE Initialize population $\mathcal{P}_0$ with $N=12$ randomly sampled HQNN architectures
\FOR{each architecture $\mathcal{A}_i \in \mathcal{P}_0$}
    \STATE Train $\mathcal{A}_i$ for $5$ epochs
    \STATE Evaluate $\mathrm{Accuracy}(\mathcal{A}_i)$, $\mathcal{E}(\mathcal{A}_i)$, and $\mathcal{G}(\mathcal{A}_i)$
\ENDFOR
\FOR{$g = 1$ to $8$}
    \STATE Perform non-dominated sorting on $\mathcal{P}_{g-1}$
    \STATE Compute crowding distance within each Pareto front
    \STATE Select parents using binary tournament selection
    \STATE Generate offspring $\mathcal{Q}_g$ using crossover and mutation
    \FOR{each offspring architecture $\mathcal{A}_j \in \mathcal{Q}_g$}
        \STATE Train $\mathcal{A}_j$ for $5$ epochs
        \STATE Evaluate $\mathrm{Accuracy}(\mathcal{A}_j)$, $\mathcal{E}(\mathcal{A}_j)$, and $\mathcal{G}(\mathcal{A}_j)$
    \ENDFOR
    \STATE Merge parent and offspring populations: $\mathcal{R}_g = \mathcal{P}_{g-1} \cup \mathcal{Q}_g$
    \STATE Perform non-dominated sorting on $\mathcal{R}_g$
    \STATE Select the next population $\mathcal{P}_g$ of size $N$ using Pareto rank and crowding distance
\ENDFOR
\STATE Return the final Pareto-optimal architectures
\end{algorithmic}
\end{algorithm}

At each generation, candidate architectures are ranked using non-dominated sorting. Parent selection is performed via binary tournament selection based on Pareto rank and crowding distance. Offspring are generated through crossover and mutation, where crossover combines architectural attributes from parent candidates and mutation introduces random variations in both quantum and classical parameters. We apply crossover to every selected parent pair, while mutation is applied independently to architectural attributes with probability $p_m=0.40$. The combined population is then re-ranked, and the next generation is formed by selecting the most competitive and diverse candidates.

%%%%%%%%%%%%%%%%%%%%%%%%%%%%%%%%%%%%%%%%%%%%%%%%%%%%%%%%%%%%%%%%%%%%%%%%%%%%%%%%%%%%%%%%%%%%%%%%%%%%%%%%%%%%%%%%%%%%%%%%%%%%%%%%%%%%%%%%%%%%%%%%%%%%%%%%%%%%%%%%%%%%%%%%%%%%%%%%%%%%%%%%%%%%%%%%%%%%%%%%%%%%%%%%%%%%%%%%%%%%

\begin{figure*}
    \centering
    \includegraphics[width=0.97\linewidth]{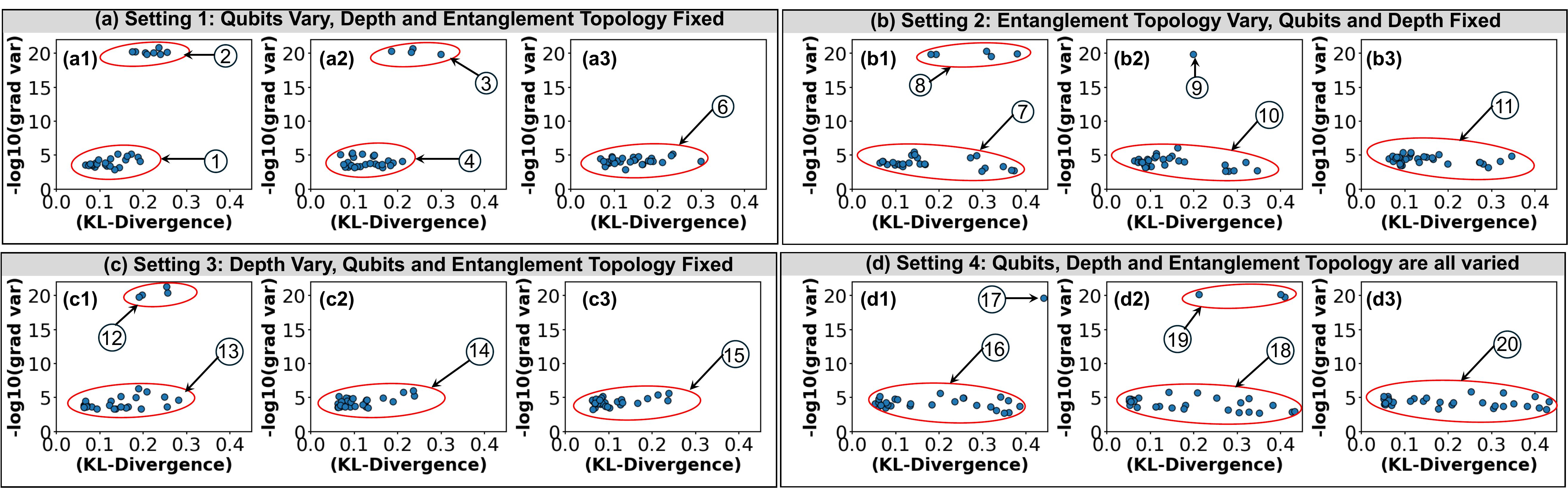}
    \caption{\textbf{Stage I} results across four architectural settings: (a) varying depth, (b) varying entanglement topology, (c) varying number of qubits, and (d) all parameters varied. Subplots (x1), (x2), and (x3) correspond to standalone PQCs, hybrid quantum-only training, and hybrid full-model training, respectively. $-log_{10} (grad var)$ is based in Eq.~\ref{eq:trainability_phase1}, where smallers values means high gradiant variance (Better trainability) and vice versa. KL-Divergence is based on Eq.~\ref{eq:expr_phase1}, where lower value means high expressibility and versa. The results show that the expressibility--trainability relationship observed in PQCs significantly weakens or breaks down in hybrid models with frozen classical layers, and is is completely eliminated under full hybrid model's end-to-end training.}
    \label{fig:res_p1}
\end{figure*}
\section{Results and Discussion}

\subsection{Stage--I Results: Empirical Analysis of expressibility-trainability in HQNNs } \label{sec:res_phase1}
In this section, we analyze the relationship between trainability, expressibility, and task performance across different architectural settings. Our goal is to examine whether the relationship between expressibility and trainability in standalone PQCs persists in hybrid models.
We consider multiple controlled settings (Table~\ref{tab:pqc_settings}) where specific architectural factors such as depth, number of qubits, and entanglement topology are varied. For each setting, we compare three configurations: pure PQC, hybrid models with full training, and hybrid models with quantum-only training.

\subsubsection{Setting 1: Fixed entanglement topology and depth, varying qubits}
The results of Setting~1 (from Table~\ref{tab:pqc_settings}) are shown in Fig.~\ref{fig:res_p1}(a). For the pure PQC case, a weak expressibility--trainability trade-off is observed, although it does not follow a smooth monotonic trend. Instead, the results form two distinct clusters. 
The first cluster corresponds to architectures with lower KL divergence, indicating higher expressibility (pointer~\rpoint{1}). In this regime, the trainability metric also remains low, reflecting relatively large gradient variance and favorable optimization behavior. In contrast, the second cluster appears at relatively higher KL divergence values, where the trainability metric increases sharply (pointer~\rpoint{2}), indicating vanishing gradients and poor trainability. This suggests a regime-dependent trade-off: trainability remains stable for moderate increases in expressibility, but degrades rapidly beyond a certain threshold.
This behavior differs slightly from prior studies, where a clearer expressibility--trainability trade-off is typically observed in standalone PQCs~\cite{Holmes_2022}. However, in such works, the trade-off is analyzed for isolated PQCs under task-independent cost functions. In our setting, the use of a task-dependent cost function (Section~\ref{sec:datasettask_phase1}) alters the optimization landscape, leading to non-monotonic and regime-dependent behavior. Furthermore, the limited qubit range ($n \in \{8,10,12\}$) results in discrete transitions rather than smooth scaling.

For the hybrid model with quantum-only gradients, the expressibility--trainability relationship becomes weaker. Although the same PQC architectures are used, fewer configurations fall into the poor-trainability regime than in the pure PQC case (pointer~\rpoint{3}), while a larger fraction of architectures remains well-trainable (pointer~\rpoint{4}). 
This behavior can be attributed to the presence of classical layers. Although classical parameters are not updated in this training configuration, they remain part of the model in the forward pass and therefore influence the cost function. Consequently, the model output is given by 
$y = g_{\boldsymbol{\phi}}\big(f_{\boldsymbol{\theta}}(h_{\boldsymbol{\phi}}(x))\big)$, as in Eq.~\ref{eq:hyb_model}, and the corresponding cost function takes the form 
$C\big(g_{\boldsymbol{\phi}}(f_{\boldsymbol{\theta}}(h_{\boldsymbol{\phi}}(x)))\big)$. 
Thus, even when gradients are computed only with respect to $\boldsymbol{\theta}$, the classical preprocessing and post-processing layers transform the input representation and reshape the effective optimization landscape. This implicit reparameterization mitigates gradient concentration effects, allowing a larger fraction of architectures to retain favorable trainability despite increases in expressibility.

In the fully trainable hybrid model, the expressibility--trainability relationship is effectively eliminated. All configurations lie within a narrow band of favorable trainability (pointer~\rpoint{5}), with no instances of the poor-trainability regime, regardless of the underlying PQC expressibility. This behavior arises from the joint training of both quantum and classical layers, which adapt during optimization, effectively reshaping the hybrid cost function $C\big(g_{\boldsymbol{\phi}}(f_{\boldsymbol{\theta}}(h_{\boldsymbol{\phi}}(x)))\big)$. This adaptive transformation acts as an implicit preconditioning mechanism, stabilizing gradient flow and preventing the onset of barren plateaus.

%%%%%%%%%%

Overall, these results demonstrate a progressive weakening of the expressibility--trainability relationship from pure PQCs to hybrid models, resulting in a complete decoupling under full hybrid training.

\subsubsection{Setting 2: Fixed Qubits and Depth, Varying Entanglement Topology}
The results of Setting~2, (from Table~\ref{tab:pqc_settings}), are shown in Fig.~\ref{fig:res_p1}(b). In the pure PQC case, the expressibility--trainability relationship becomes less structured compared to Setting~1. Although two trainability regimes are still visible, namely a well-trainable region with lower values of $-\log_{10}(\mathrm{grad\;var})$ (pointer~\rpoint{7}) and a poorly trainable region with higher values (pointer~\rpoint{8}), these regimes are no longer clearly separated by expressibility. Instead, both well- and poorly-trainable configurations appear across a broad range of KL-divergence values.
This suggests that, when the number of qubits and depth are fixed, expressibility alone is insufficient to explain trainability. Since the Hilbert-space dimension is fixed in this setting, changing the entanglement topology mainly alters how correlations are distributed across qubits rather than changing the representational dimension of the circuit. As a result, different entanglement patterns can achieve similar levels of expressibility while inducing substantially different gradient-variance behavior. Therefore, the expressibility--trainability trade-off becomes weaker and more topology-dependent than in settings where the number of qubits is varied.

Consistent with the trends discussed in Setting~1, the hybrid model with quantum-only gradients further weakens this relationship, with most configurations shifting toward favorable trainability (pointer~\rpoint{10}) and very few remaining poorly trainable (pointer~\rpoint{9}). 
In the fully hybrid setting, the relationship is effectively eliminated, with all configurations exhibiting stable trainability (pointer~\rpoint{11}), indicating that classical layers compensate for entanglement topology-induced variations.

%%%%%%%%%%%%%%%%%%
\subsubsection{Setting 3: Fixed Qubits and Entanglement Topology, Varying Depth}

The results of Setting~3 are shown in Fig.~\ref{fig:res_p1}(c). In the pure PQC case, a weak and non-monotonic expressibility--trainability relationship is observed as depth increases. While higher expressibility occasionally correlates with degraded trainability (pointer~\rpoint{12}), several deeper circuits remain well-trainable (pointer~\rpoint{13}). This indicates a regime-dependent behavior rather than a strict trade-off, consistent with the fact that increasing depth does not expand the Hilbert space, and therefore has a comparatively limited impact on expressibility relative to increasing the number of qubits.

Consistent with earlier settings, both hybrid variants eliminate the poorly trainable regime. In the quantum-only hybrid case, all configurations remain well-trainable (pointer~\rpoint{14}), indicating that classical layers are sufficient to suppress depth-induced gradient concentration. In the fully hybrid setting, this behavior persists (pointer~\rpoint{15}), confirming that the expressibility--trainability relationship is effectively removed under hybrid architectures in this setting.

\begin{figure*}[b]
    \centering
    \includegraphics[width=0.98\linewidth]{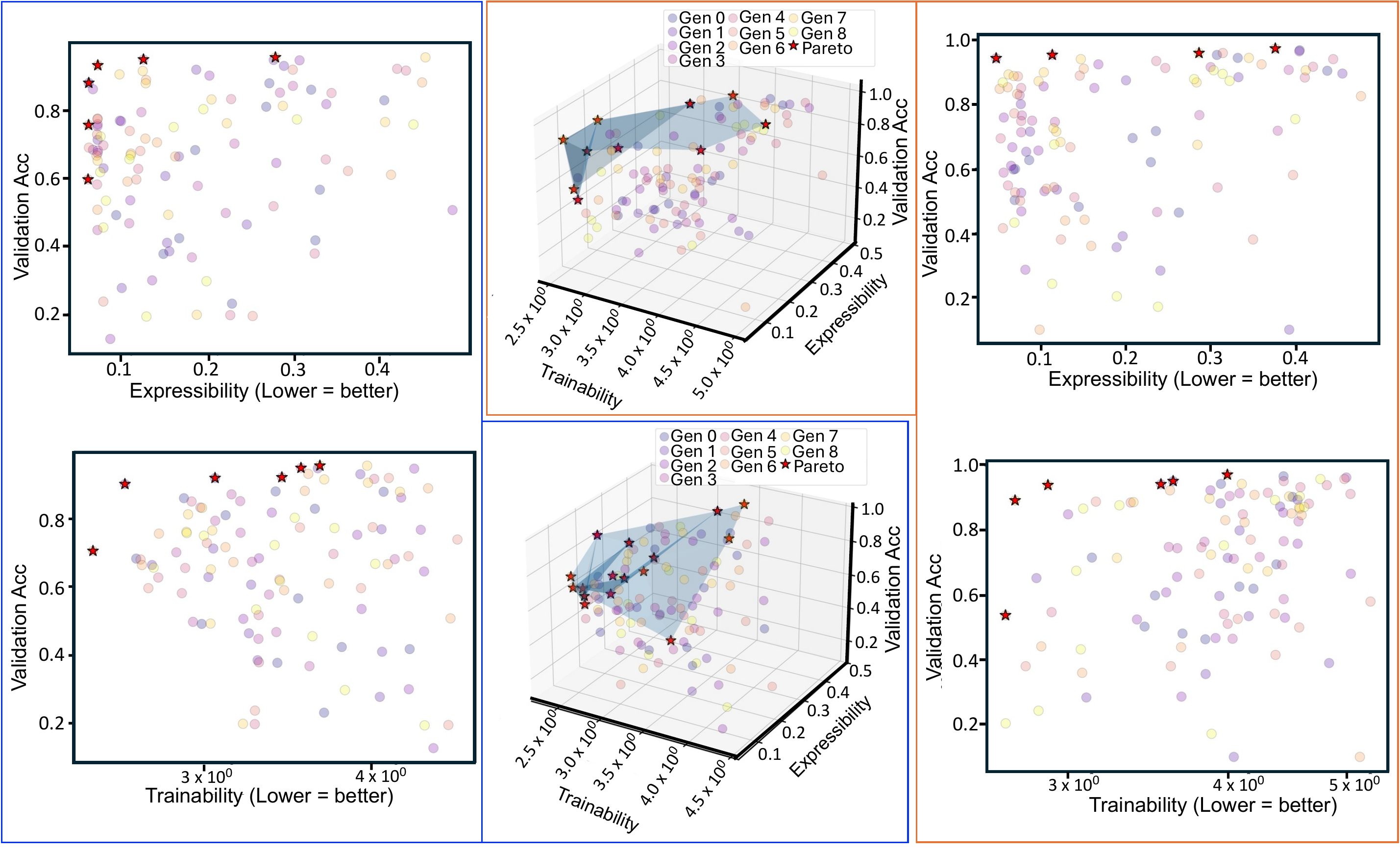}
    \caption{NAS results for HQNN architectures under two trainability definitions. The blue panels correspond to full-model trainability (gradients computed with respect to all model parameters), while the orange panels correspond to quantum-only trainability (gradients computed with respect to quantum parameters only in hybrid setting). Each point represents a candidate architecture, and red stars indicate Pareto-optimal solutions. The plots show the trade-offs between validation accuracy, expressibility, and trainability.}
    \label{fig:res_nas_full}
\end{figure*}

\subsubsection{Setting 4: Fully Variable Architecture}

The results of Setting~4 are shown in Fig.~\ref{fig:res_p1}(d), where all architectural components (qubits, depth, and entanglement topology) are varied jointly. In the pure PQC case, the expressibility--trainability relationship becomes highly heterogeneous. Most configurations lie within a well-trainable region (pointer~\rpoint{16}), while a small number of outliers exhibit significantly degraded trainability (pointer~\rpoint{17}).
Unlike previous settings, no clear structure or regime transition is observed with respect to expressibility. Architectures with similar KL-divergence can exhibit markedly different trainability, indicating that expressibility alone is insufficient to predict optimization behavior in the full design space. Instead, trainability depends on the combined interaction between qubit count, depth, and entanglement topology, making the relationship inherently architecture-dependent. This further motivates the need for quantum architecture search, even in pure PQCs.

In the hybrid model with quantum-only gradients, this heterogeneity persists, and in contrast to earlier settings, the number of poorly trainable configurations slightly increases. While most architectures remain well-trainable (pointer~\rpoint{18}), additional configurations appear in the degraded regime (pointer~\rpoint{19}). This can be attributed to the fixed nonlinear transformation induced by the frozen classical layers, which affects different architectures unevenly and redistributes gradient statistics across the design space.
In the fully hybrid setting, however, all configurations remain well-trainable (pointer~\rpoint{20}), confirming that end-to-end training effectively compensates for architecture-dependent variations and eliminates the expressibility--trainability trade-off.
%%%%%%%%%%%%%%%%%%%%%%%%%%%%%%%%%%%%%%%%%%%%%%%%%%%%%%%%%%%%%%%%%%%%%%%%%%%%%%%%%%%%%%

\subsection{Stage--II Results: Neural Architecture Search} \label{sec:res_phase2}

Building on the observations from Section~\ref{sec:res_phase1}, where no consistent expressibility--trainability relationship was observed in HQNNs, we perform multi-objective neural architecture search (NAS) to identify Pareto-optimal hybrid architectures.
Unlike state-of-the-art quantum architecture search, which focuses solely on optimizing the quantum circuit~\cite{li2023eqnas,duong2022,kashif2025faqnas} we consider a joint classical--quantum search space. This design choice is directly motivated by the findings in Section~\ref{sec:res_phase1}, where classical components were shown to significantly influence the optimization landscape. Consequently, restricting the search to the quantum circuit alone can lead to suboptimal or misleading conclusions, whereas joint search enables a more complete and faithful exploration of hybrid model design.
As outlined in Section~\ref{sec:NAS_objectives}, we consider two notions of trainability in hybrid models: (i) \textit{full-model trainability}, where both classical and quantum parameters are optimized end-to-end, and (ii) \textit{quantum-only trainability}, where only the quantum parameters are updated while the classical layers remain fixed during training.

\begin{table*}
\centering
\footnotesize
\caption{Specifications of Pareto-optimal architectures obtained from NAS using full hybrid model trainability.}
\label{tab:pareto_full_trainability}
\resizebox{\textwidth}{!}{
\begin{tabular}{c l c c c c c c}
\toprule
\rowcolor{gray!15}
\textbf{Qubits} & \textbf{Parameterized Gates} & \textbf{Ent. Topology} & \textbf{Ent. Gate} & \textbf{Depth} & \textbf{Accuracy} & \textbf{Trainability} & \textbf{Expressibility} \\
\midrule

12 & RY-RX-RX-RZ-RX-RY-RZ-RZ-RY-RZ-RX-RZ & all-to-all & CZ & 45 & 0.5963 & 3.7456 & 0.0499 \\

9 & RX-RX-RZ-RY-RY-RX-RY-RZ-RZ & circular & CZ & 50 & 0.5963 & 2.5803 & 0.1059 \\

9 & RY-RX-RX-RY-RZ-RY-RZ-RZ-RY & circular & CZ & 45 & 0.6563 & 2.6039 & 0.0979 \\

10 & RX-RY-RX-RZ-RY-RX-RZ-RZ-RX-RZ & paired & CZ & 35 & 0.6800 & 2.5453 & 0.1105 \\

9 & RX-RX-RZ-RZ-RY-RX-RX-RZ-RY & paired & CZ & 50 & 0.7050 & 2.3417 & 0.1299 \\

12 & RX-RY-RX-RZ-RY-RY-RX-RZ-RY-RZ-RX-RZ & all-to-all & CZ & 45 & 0.7563 & 2.6287 & 0.0505 \\

10 & RX-RZ-RY-RZ-RY-RY-RZ-RY-RY-RZ & all-to-all & CNOT & 50 & 0.7588 & 3.0324 & 0.0609 \\

9 & RX-RX-RX-RZ-RY-RX-RX-RZ-RY & circular & CZ & 50 & 0.7938 & 2.9004 & 0.1118 \\

12 & RY-RX-RX-RX-RX-RX-RY-RZ-RY-RX-RX-RZ & all-to-all & CZ & 45 & 0.8800 & 3.2248 & 0.0508 \\

10 & RX-RX-RX-RZ-RY-RX-RX-RZ-RY-RZ & paired & CNOT & 50 & 0.9013 & 2.4762 & 0.1906 \\

10 & RX-RX-RY-RZ-RX-RX-RX-RY-RY-RZ & paired & CNOT & 45 & 0.9188 & 2.8999 & 0.1752 \\

10 & RY-RZ-RX-RZ-RX-RX-RX-RZ-RX-RZ & alternating & CZ & 15 & 0.9213 & 3.2606 & 0.4082 \\

10 & RX-RX-RX-RZ-RY-RX-RX-RZ-RY-RZ & all-to-all & CNOT & 20 & 0.9313 & 3.4140 & 0.0611 \\

9 & RX-RX-RX-RZ-RY-RX-RY-RZ-RX & circular & CZ & 50 & 0.9488 & 3.3706 & 0.1141 \\

10 & RY-RZ-RX-RZ-RX-RX-RX-RZ-RX-RZ & paired & CZ & 5 & 0.9550 & 3.4852 & 0.4411 \\

9 & RX-RY-RX-RZ-RY-RX-RY-RZ-RX & circular & CZ & 30 & 0.9550 & 3.7853 & 0.2671 \\

\bottomrule
\end{tabular}}
\end{table*}

\paragraph{NAS Results: Full Hybrid Model Trainability}

The NAS results when trainability is defined using full-model gradient variance are shown in Fig.~\ref{fig:res_nas_full} (enclosed in blue). We identify Pareto-optimal architectures with respect to accuracy, expressibility, and trainability, and analyze the trade-offs captured by the resulting Pareto front.
We observe that no single objective dominates performance. High accuracy can be achieved across a range of expressibility and trainability values, indicating that neither expressibility nor trainability alone is sufficient to characterize optimal hybrid architectures. In particular, architectures with moderate expressibility tend to achieve the best performance, while highly expressive configurations do not consistently lead to improved accuracy. This may be attributed to the increased complexity of the optimization landscape in highly expressive circuits, as well as the ability of classical components in hybrid models which can potentially compensate for suboptimal quantum optimization.

The Pareto front highlights clear trade-offs between the three objectives. Some architectures achieve strong trainability at the cost of reduced accuracy, while others prioritize accuracy with less favorable optimization properties. A subset of architectures provides a balanced trade-off, achieving competitive accuracy with moderate expressibility and good trainability. This behavior reflects the interplay between representation capacity and optimization efficiency in hybrid models. Representative Pareto-optimal architectures in Table~\ref{tab:pareto_full_trainability} further illustrate these trade-offs, showing that high-performing models span diverse configurations in terms of circuit depth, qubit count, and entanglement structure.
Importantly, compared to the analysis in Section~\ref{sec:res_phase1}, trainability now exhibits a more meaningful relationship with performance, suggesting that joint classical-quantum architecture search restores a degree of structure to the optimization landscape. However, this relationship remains non-trivial, reinforcing that HQNN design cannot be guided by a single metric.

Overall, these results demonstrate that effective HQNN architectures emerge from balancing competing objectives, rather than optimizing expressibility or trainability of quantum circuits in isolation.

\begin{table*}[t!]
\centering
\footnotesize
\caption{Specifications of Pareto-optimal architectures obtained from NAS using quantum-only trainability in hybrid model.}
\label{tab:pareto_quantum_trainability}
\resizebox{\textwidth}{!}{
\begin{tabular}{c l c c c c c c}
\toprule
\rowcolor{gray!15}
\textbf{Qubits} & \textbf{Parameterized Gates} & \textbf{Ent. Topology} & \textbf{Ent. Gate} & \textbf{Depth} & \textbf{Accuracy} & \textbf{Trainability} & \textbf{Expressibility} \\
\midrule
10 & RX-RY-RZ-RX-RX-RX-RZ-RZ-RY-RX & circular & CNOT & 4 & 0.5375 & 2.5492 & 0.1059 \\
9  & RX-RY-RZ-RZ-RY-RZ-RX-RZ-RZ & star & CZ & 10 & 0.5475 & 2.7734 & 0.0604 \\
12 & RX-RY-RZ-RX-RX-RZ-RZ-RY-RZ-RX-RX-RX & star & CZ & 10 & 0.8675 & 2.9396 & 0.0542 \\
11 & RY-RY-RX-RZ-RY-RY-RX-RZ-RZ-RX-RX & star & CZ & 10 & 0.8925 & 2.5941 & 0.0610 \\
9  & RX-RX-RY-RY-RZ-RY-RZ-RX-RX & paired & CZ & 10 & 0.9400 & 2.7551 & 0.1465 \\
12 & RX-RY-RZ-RX-RX-RZ-RX-RZ-RZ-RX-RZ-RX & star & CZ & 20 & 0.9425 & 3.3896 & 0.0465 \\
10 & RZ-RX-RZ-RZ-RY-RX-RX-RX-RY-RY & alternating & CZ & 4  & 0.9525 & 3.4655 & 0.3070 \\
11 & RX-RX-RZ-RY-RX-RZ-RZ-RX-RX-RX-RY & alternating & CZ & 45 & 0.9588 & 4.5404 & 0.2859 \\
11 & RX-RX-RZ-RX-RX-RZ-RZ-RZ-RX-RX-RY & paired & CZ & 50 & 0.9525 & 4.2443 & 0.1127 \\

11 & RZ-RX-RZ-RX-RX-RZ-RY-RX-RY-RZ-RX & star & CZ & 10 & 0.9725 & 3.8281 & 0.3760 \\

\bottomrule
\end{tabular}}
\end{table*}
\paragraph{NAS Results: Quantum-Only Trainability in Hybrid Models}

Fig.~\ref{fig:res_nas_full} (enclosed in orange), shows the NAS results when trainability is defined using quantum-only gradient variance. Similar to the full-model case, we observe a well-defined Pareto front, indicating clear trade-offs between accuracy, expressibility, and trainability.
Compared to full-model training, the distribution of architectures is slightly more structured. This suggests that removing classical gradient contributions reduces variability in the optimization landscape and makes the influence of quantum circuit properties more evident.
In particular, architectures with moderate expressibility tend to achieve the best performance, while highly expressive configurations do not consistently yield improved accuracy. This trend is more visible compared to the full-model case, indicating that classical components may obscure underlying relationships between quantum properties and performance.

The Pareto front highlights these trade-offs, with architectures achieving different balances between performance and optimization properties. Representative Pareto-optimal architectures in Table~\ref{tab:pareto_quantum_trainability} further illustrate this behavior, showing that high-performing models span diverse quantum circuit configurations.
However, trainability still does not exhibit a clear monotonic relationship with accuracy, and high-performing architectures are observed across a range of trainability values. This indicates that, even when isolating quantum gradients, no single objective provides sufficient guidance for architecture selection.

Overall, these results reinforce that HQNN design remains inherently multi-objective, and that while isolating quantum trainability reveals partial structure, it does not fully recover the behavior observed in standalone PQCs.

%%%%%%%%%%%%%%%%%%%%%%%%%%%%%%%%%%%%%%%%%%%%%%%%%%%%%%%%%%%%%%%%%%%%%%%%%%%%%%%%%%%%%%%%%%%%%%%%%%%%%%%%%%%%%%%%%%%%%%%%%%%%%%%%%%%%%%

\section{Conclusion}

We investigated the expressibility--trainability relationship in hybrid quantum neural networks (HQNNs) and its implications for architecture design. Our results show that the commonly assumed trade-off, well-established for standalone PQCs, holds only weakly under task-specific settings, where deviations arise from the choice of cost function.
In hybrid models, this relationship is significantly disrupted. Even with frozen classical layers, classical pre-processing alters the effective input space, reshaping the optimization landscape and weakening the dependence of trainability on circuit expressibility. When classical and quantum parameters are trained jointly, this relationship effectively disappears, with no consistent correspondence between expressibility and trainability of the underlying quantum layers.
These findings indicate that the expressibility--trainability trade-off is not a reliable design principle for HQNNs. 

Motivated by this, we formulated HQNN architecture design as a multi-objective optimization problem and employed neural architecture search (NAS) to jointly optimize expressibility, trainability, and task performance over a combined classical--quantum design space. Our results show that effective architectures emerge from balancing these competing objectives, rather than optimizing any single metric in isolation.
Overall, this work establishes that HQNN design is inherently multi-objective and fundamentally shaped by classical--quantum interactions, challenging PQC-centric design assumptions. It provides a foundation for systematic, data-driven exploration of hybrid architectures and opens new directions for principled design methodologies in quantum machine learning.

\section*{Acknowledgment}
This work was supported in part by the NYUAD Center for Quantum and Topological Systems (CQTS), funded by Tamkeen under the NYUAD Research Institute grant CG008.

\bibliographystyle{IEEEtran}
\bibliography{main}

@article{kashif2026closing,
  title={Closing the Loop: Resource-aware Hybrid NAS Guided by Analytical and Hardware-Calibrated Quantum Cost Modeling},
  author={Kashif, Muhammad and Marchisio, Alberto and Shafique, Muhammad},
  journal={arXiv preprint arXiv:2603.00625},
  year={2026}
}

@article{marchisio2026hybrid,
  title={Hybrid Quantum-Classical Neural Architecture Search},
  author={Marchisio, Alberto and Kashif, Muhammad and Innan, Nouhaila and Shafique, Muhammad},
  journal={arXiv preprint arXiv:2605.18345},
  year={2026}
}

@inproceedings{kashif2025computational,
  title={Computational advantage in hybrid quantum neural networks: Myth or reality?},
  author={Kashif, Muhammad and Marchisio, Alberto and Shafique, Muhammad},
  booktitle={2025 62nd ACM/IEEE Design Automation Conference (DAC)},
  pages={1--7},
  year={2025},
  organization={IEEE}
}

@article{duong2022,
      title={Quantum Neural Architecture Search with Quantum Circuits Metric and Bayesian Optimization}, 
      author={Trong Duong and Sang T. Truong and Minh Tam and Bao Bach and Ju-Young Ryu and June-Koo Kevin Rhee},
      year={2022},
      eprint={2206.14115},
      journal={arXiv:2206.14115},
      primaryClass={quant-ph},
      url={https://arxiv.org/abs/2206.14115}, 
}

@article{kashif2025faqnas,
  title={FAQNAS: FLOPs-aware Hybrid Quantum Neural Architecture Search using Genetic Algorithm},
  author={Kashif, Muhammad and Khalid, Shaf and Marchisio, Alberto and Innan, Nouhaila and Shafique, Muhammad},
  journal={arXiv preprint arXiv:2511.10062},
  year={2025}
}

@article{li2023eqnas,
  title={EQNAS: Evolutionary quantum neural architecture search for image classification},
  author={Li, Yangyang and Liu, Ruijiao and Hao, Xiaobin and Shang, Ronghua and Zhao, Peixiang and Jiao, Licheng},
  journal={Neural Networks},
  volume={168},
  pages={471--483},
  year={2023},
  publisher={Elsevier}
}

@article{hubregtsen2021evaluation,
  title={Evaluation of parameterized quantum circuits: on the relation between classification accuracy, expressibility, and entangling capability},
  author={Hubregtsen, Thomas and Pichlmeier, Josef and Stecher, Patrick and Bertels, Koen},
  journal={Quantum Machine Intelligence},
  volume={3},
  number={1},
  pages={9},
  year={2021},
  publisher={Springer}
}

@article{zhang2025learning,
  title={Learning the expressibility of quantum circuit ansatz using transformer},
  author={Zhang, Fei and Li, Jie and He, Zhimin and Situ, Haozhen},
  journal={Advanced Quantum Technologies},
  volume={8},
  number={6},
  pages={2400366},
  year={2025},
  publisher={Wiley Online Library}
}

@article{cerezo2021variational,
  title={Variational quantum algorithms},
  author={Cerezo, Marco and Arrasmith, Andrew and Babbush, Ryan and Benjamin, Simon C and Endo, Suguru and Fujii, Keisuke and McClean, Jarrod R and Mitarai, Kosuke and Yuan, Xiao and Cincio, Lukasz and others},
  journal={Nature Reviews Physics},
  volume={3},
  number={9},
  pages={625--644},
  year={2021},
  publisher={Nature Publishing Group UK London}
}

@article{amaro2022filtering,
  title={Filtering variational quantum algorithms for combinatorial optimization},
  author={Amaro, David and Modica, Carlo and Rosenkranz, Matthias and Fiorentini, Mattia and Benedetti, Marcello and Lubasch, Michael},
  journal={Quantum Science \& Technology},
  volume={7},
  number={1},
  pages={015021},
  year={2022},
  publisher={IOP Publishing}
}

@inproceedings{innan2025quav,
  title={QUAV: Quantum-Assisted Path Planning and Optimization for UAV Navigation with Obstacle Avoidance},
  author={Innan, Nouhaila and Kashif, Muhammad and Marchisio, Alberto and Gan, Yung-Sze and Barbaresco, Frederic and Shafique, Muhammad},
  booktitle={2025 IEEE International Conference on Quantum Artificial Intelligence (QAI)},
  pages={208--215},
  year={2025},
  organization={IEEE}
}

@inproceedings{kashif2025position,
  title={Position Paper: Quantum Neural Networks-A Paradigm Shift in AI or a Theoretical Promise?},
  author={Kashif, Muhammad and Shafique, Muhammad},
  booktitle={2025 International Joint Conference on Neural Networks (IJCNN)},
  pages={1--10},
  year={2025},
  organization={IEEE}
}

@article{schuld2015introduction,
  title={An introduction to quantum machine learning},
  author={Schuld, Maria and Sinayskiy, Ilya and Petruccione, Francesco},
  journal={Contemporary Physics},
  volume={56},
  number={2},
  pages={172--185},
  year={2015},
  publisher={Taylor \& Francis}
}

@article{Sim_2019,
   title={Expressibility and Entangling Capability of Parameterized Quantum Circuits for Hybrid Quantum‐Classical Algorithms},
   volume={2},
   ISSN={2511-9044},
   url={http://dx.doi.org/10.1002/qute.201900070},
   DOI={10.1002/qute.201900070},
   number={12},
   journal={Advanced Quantum Technologies},
   publisher={Wiley},
   author={Sim, Sukin and Johnson, Peter D. and Aspuru‐Guzik, Alán},
   year={2019},
   month=oct }

@article{heyraud2023efficient,
  title={Efficient estimation of trainability for variational quantum circuits},
  author={Heyraud, Valentin and Li, Zejian and Donatella, Kaelan and Le Boit{\'e}, Alexandre and Ciuti, Cristiano},
  journal={PRX Quantum},
  volume={4},
  number={4},
  pages={040335},
  year={2023},
  publisher={APS}
}

@article{mcclean2018barren,
  title={Barren plateaus in quantum neural network training landscapes},
  author={McClean, Jarrod R and Boixo, Sergio and Smelyanskiy, Vadim N and Babbush, Ryan and Neven, Hartmut},
  journal={Nature communications},
  volume={9},
  number={1},
  pages={4812},
  year={2018},
  publisher={Nature Publishing Group UK London}
}

@article{Holmes_2022,
   title={Connecting Ansatz Expressibility to Gradient Magnitudes and Barren Plateaus},
   volume={3},
   ISSN={2691-3399},
   url={http://dx.doi.org/10.1103/PRXQuantum.3.010313},
   DOI={10.1103/prxquantum.3.010313},
   number={1},
   journal={PRX Quantum},
   publisher={American Physical Society (APS)},
   author={Holmes, Zoë and Sharma, Kunal and Cerezo, M. and Coles, Patrick J.},
   year={2022},
   month=jan }

@article{kashif2024resqnets,
  title={Resqnets: a residual approach for mitigating barren plateaus in quantum neural networks},
  author={Kashif, Muhammad and Al-Kuwari, Saif},
  journal={EPJ Quantum Technology},
  volume={11},
  number={1},
  pages={1--28},
  year={2024},
  publisher={Springer}
}

@inproceedings{kashif2024alleviating,
  title={Alleviating barren plateaus in parameterized quantum machine learning circuits: Investigating advanced parameter initialization strategies},
  author={Kashif, Muhammad and Rashid, Muhammad and Al-Kuwari, Saif and Shafique, Muhammad},
  booktitle={2024 Design, Automation \& Test in Europe Conference \& Exhibition (DATE)},
  pages={1--6},
  year={2024},
  organization={IEEE}
}

@article{tangpanitanon2020,
  title={Expressibility and trainability of parametrized analog quantum systems for machine learning applications},
  author={Tangpanitanon, Jirawat and Thanasilp, Supanut and Dangniam, Ninnat and Lemonde, Marc-Antoine and Angelakis, Dimitris G},
  journal={Physical Review Research},
  volume={2},
  number={4},
  pages={043364},
  year={2020},
  publisher={APS}
}

@article{roseler2026find,
  title={How to find expressible and trainable parameterized quantum circuits?},
  author={R{\"o}seler, Peter and Willsch, Dennis and Michielsen, Kristel},
  journal={arXiv preprint arXiv:2603.14451},
  year={2026}
}

@article{maronese2026high,
  title={High-expressibility quantum neural networks using only classical resources},
  author={Maronese, Marco and Ferrari, Francesco and Vandelli, Matteo and Dragoni, Daniele},
  journal={Quantum Machine Intelligence},
  volume={8},
  number={1},
  pages={25},
  year={2026},
  publisher={Springer}
}

@article{kashif2023unified,
  title={The unified effect of data encoding, ansatz expressibility and entanglement on the trainability of hqnns},
  author={Kashif, Muhammad and Al-Kuwari, Saif},
  journal={International Journal of Parallel, Emergent and Distributed Systems},
  volume={38},
  number={5},
  pages={362--400},
  year={2023},
  publisher={Taylor \& Francis}
}

@article{sanz2026efficiently,
  title={Efficiently architecting VQAs: Expressibility--Trainability--Resources Pareto-Optimality},
  author={Sanz, Rodrigo M and Angles-Castillo, Andreu and Alarcon, Eduard and Almudever, Carmen G},
  journal={arXiv preprint arXiv:2603.22142},
  year={2026}
}

@ARTICLE{yu:2025,
  author={Liu, Yu and Kaneko, Kazuya and Baba, Kentaro and Koyama, Jumpei and Kimura, Koichi and Takeda, Naoyuki},
  journal={IEEE Transactions on Quantum Engineering}, 
  title={Analysis of Parameterized Quantum Circuits: On the Connection Between Expressibility and Types of Quantum Gates}, 
  year={2025},
  volume={6},
  number={},
  pages={1-12},
  doi={10.1109/TQE.2025.3571484}}

@article{thanasilp2023,
  title={Subtleties in the trainability of quantum machine learning models},
  author={Thanasilp, Supanut and Wang, Samson and Nghiem, Nhat Anh and Coles, Patrick and Cerezo, Marco},
  journal={Quantum Machine Intelligence},
  volume={5},
  number={1},
  pages={21},
  year={2023},
  publisher={Springer}
}

@article{kashif2021design,
  title={Design space exploration of hybrid quantum--classical neural networks},
  author={Kashif, Muhammad and Al-Kuwari, Saif},
  journal={Electronics},
  volume={10},
  number={23},
  pages={2980},
  year={2021},
  publisher={MDPI}
}

@article{paquet2022quantumleap,
  title={QuantumLeap: Hybrid quantum neural network for financial predictions},
  author={Paquet, Eric and Soleymani, Farzan},
  journal={Expert Systems with Applications},
  volume={195},
  pages={116583},
  year={2022},
  publisher={Elsevier}
}

@article{sagingalieva2023hybrid,
  title={Hybrid quantum neural network for drug response prediction},
  author={Sagingalieva, Asel and Kordzanganeh, Mohammad and Kenbayev, Nurbolat and Kosichkina, Daria and Tomashuk, Tatiana and Melnikov, Alexey},
  journal={Cancers},
  volume={15},
  number={10},
  pages={2705},
  year={2023},
  publisher={MDPI}
}

@article{kashif2026design,
  title={Design Space Exploration of Hybrid Quantum Neural Networks for Chronic Kidney Disease},
  author={Kashif, Muhammad and Siraj, Hanzalah Mohamed and Innan, Nouhaila and Marchisio, Alberto and Shafique, Muhammad},
  journal={arXiv preprint arXiv:2604.13608},
  year={2026}
}

@article{liu2021hybrid,
  title={Hybrid quantum-classical convolutional neural networks},
  author={Liu, Junhua and Lim, Kwan Hui and Wood, Kristin L and Huang, Wei and Guo, Chu and Huang, He-Liang},
  journal={Science China Physics, Mechanics \& Astronomy},
  volume={64},
  number={9},
  pages={290311},
  year={2021},
  publisher={Springer}
}

@article{wierichs2022general,
  title={General parameter-shift rules for quantum gradients},
  author={Wierichs, David and Izaac, Josh and Wang, Cody and Lin, Cedric Yen-Yu},
  journal={Quantum},
  volume={6},
  pages={677},
  year={2022},
  publisher={Verein zur F{\"o}rderung des Open Access Publizierens in den Quantenwissenschaften}
}

@article{kashif2023impact,
  title={The impact of cost function globality and locality in hybrid quantum neural networks on nisq devices},
  author={Kashif, Muhammad and Al-Kuwari, Saif},
  journal={Machine Learning: Science and Technology},
  volume={4},
  number={1},
  pages={015004},
  year={2023},
  publisher={IOP Publishing}
}

% \end{spacing}
\end{document}